**Fat to Muscle Ratio Measurements with Dual Energy X Ray Absorbtiometry**

A. Chen[a], ,A. Wang[b], C. Broadbent[b], J. Zhong[b], A. Dilmanian[c], F. Zafonte[c], and Z. Zhong[c*]

a. Shenzhen College of International Education, 1st HuangGang Park St., Shenzhen, GuangDong, China

b. Wald Melville High school, 380 Old Town Rd., East Setauket, NY 11733, USA

c. Dept. of Radiology, Stony Brook University, Stony Brook, NY 11794, UA

**d.** National Synchrotron Light Source, Bldg. 743, Brookhaven National Laboratory, Upton, NY 11973, USA

* corresponding author

**Abstract**

Accurate measurement of the fat-to-muscle ratio in animal model is important for obesity research. An efficient way to measure the fat to muscle ratio in animal model using dual-energy absorptiometry is presented in this paper. A radioactive source exciting x-ray fluorescence from a target material is used to provide the two x-ray energies needed. The x-rays, after transmitting through the sample, are measured with an energy-sensitive Ge detector. Phantoms and specimens were measured. The results showed that the method was sensitive to the fat to muscle ratios with good linearity. A standard deviation of a few percent in the fat to muscle ratio could be observed with the x-ray dose of 0.001 mGy.

**Introduction**

There is a need for accurate and cost effective measurement of fat to muscle ratio in the testing of diabetic weight control drugs and obesity research. Currently micro-CT scans are used to measure the fat to muscle ratios of the animal models, as weighing the animals highly inaccurate due to outside variables [1 . However, one problem with these micro-CT scans is the expenses and time that is needed to obtain results on a single mouse; about an hour itself, and a few more hours when taking into account that the data obtained must be analyzed and processed. These animals need to be anesthetized, and it would take a lot of time and effort to do this for each animal considering animal experiments typically involve a large number of them. Thus micro-CT scans are highly inefficient.

A more efficient, and quicker way to measure the fat to muscle ratio is presented in this paper. X-ray attentuation depends upon the x-ray energy, and this dependency differs between fat and muscle. Thus by taking x-ray attenuation measurements of a specimen at two x-ray energies, the mass of each can be solved. This method is called dual energy x-ray absorptiometry. Dual energy x-ray absorptiometry has been widely used for differentiating between high-Z and low-Z materials, such as between bone and soft tissue for bone densitometry [2], and recently Iodine and soft tissue for transvenous coronary angiography [3-5]. Here, the use of dual energy x-ray absorptiometry is used to differentiate between two soft tissue types, the fat and muscle, and to measure fat to muscle ratio.

Using a commercially available radioactive source that emits two x-ray energies, and an x-ray detector that can discriminate between the energies, the dual energy x-ray absorptiometry can be easily performed. An experimental apparatus was designed and constructed using a radioactive source made by Amersham which provides the 22.105 and 59.5 keV x-rays and a Germanium detector made by Canberra.

Phantoms of plexiglass, which are similar to fat in x-ray absorbtion, and water, which is similar to muscle, were designed and constructed. By varying the thickness, measurements were made to determine the appropriate dosage to obtain accurate data. The amount of each component present was also changed to test the methods sensitivity and linearity.

Specimens of pork fat and muscle with different fat to muscle ratios were measured to validate this method's applicability for animal research as well. By using and layering fat and muscle, a condition that is similar to the testing of live animals is created.

**Theory**

As x-rays travels through materials, they are attenuated by the material. X-ray attenuation is described by the Beer's law

$$I = I_0 \exp(-\mu T) \qquad \text{Eq. 1}$$

where $I_0$ is the x-ray intensity before the material, $I$ is the x-ray intensity after the material, $\mu$ is the linear attenuation coefficient of the material, and $T$ is the thickness of the material.

Figure 1 shows x-rays transmitting a sample consisting of fat and muscle. The x-ray intensity after transmitting the material is

$$I = I_0 \exp(-\mu_f T_f) \exp(-\mu_m T_m) \qquad \text{Eq. 2}$$

where $\mu_f$ and $\mu_m$ are the linear attenuation coefficients for fat and muscle, respectively, and $T_f$ and $T_m$ are the thicknesses of fat and muscle, respectively.

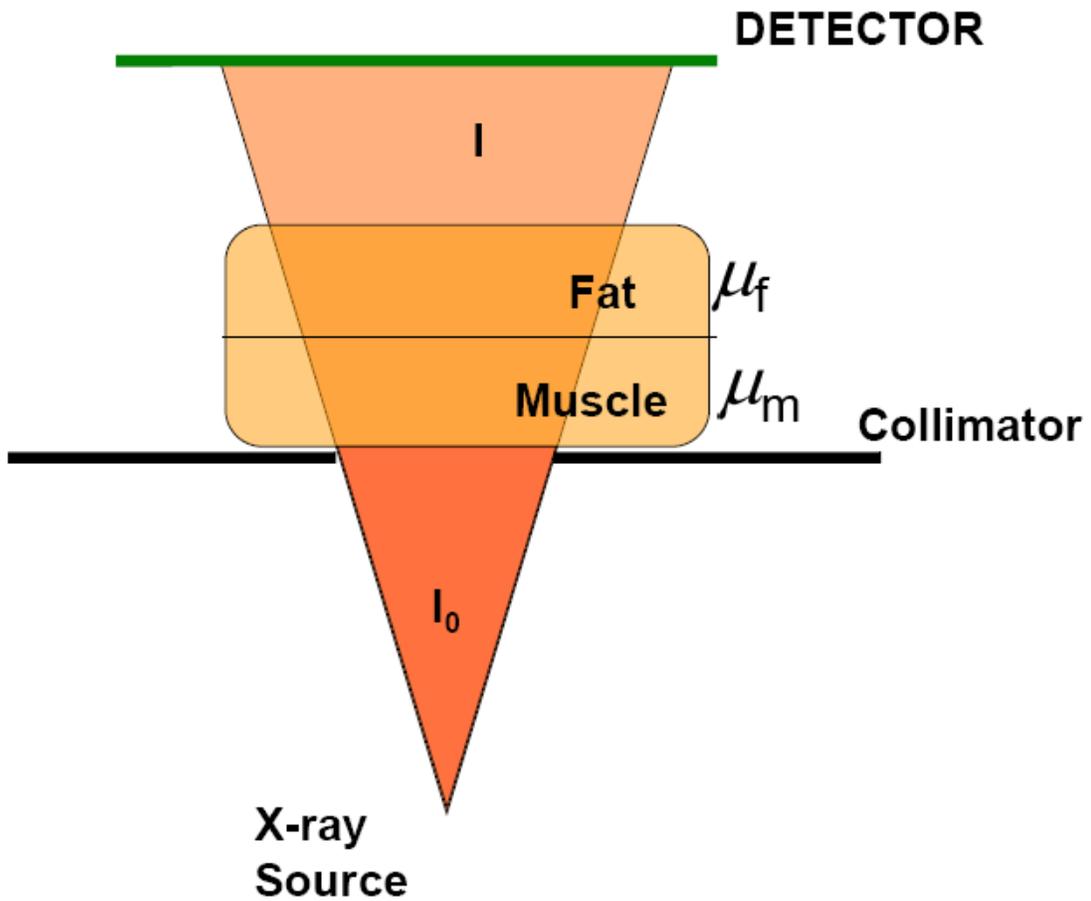

**Figure 1. This was the design used to calculate the ratios. The X-ray source is placed under the detector, with the specimen in between.**

In Eq. 1, the linear attenuation coefficient $\mu$ depends on the x-ray energy. Higher the x-ray energy corresponds to lower attenuation coefficient. Figure 1 shows the theoretial dependence of the linear attenuation coefficients of fat and muscle as a function of x-ray energy.

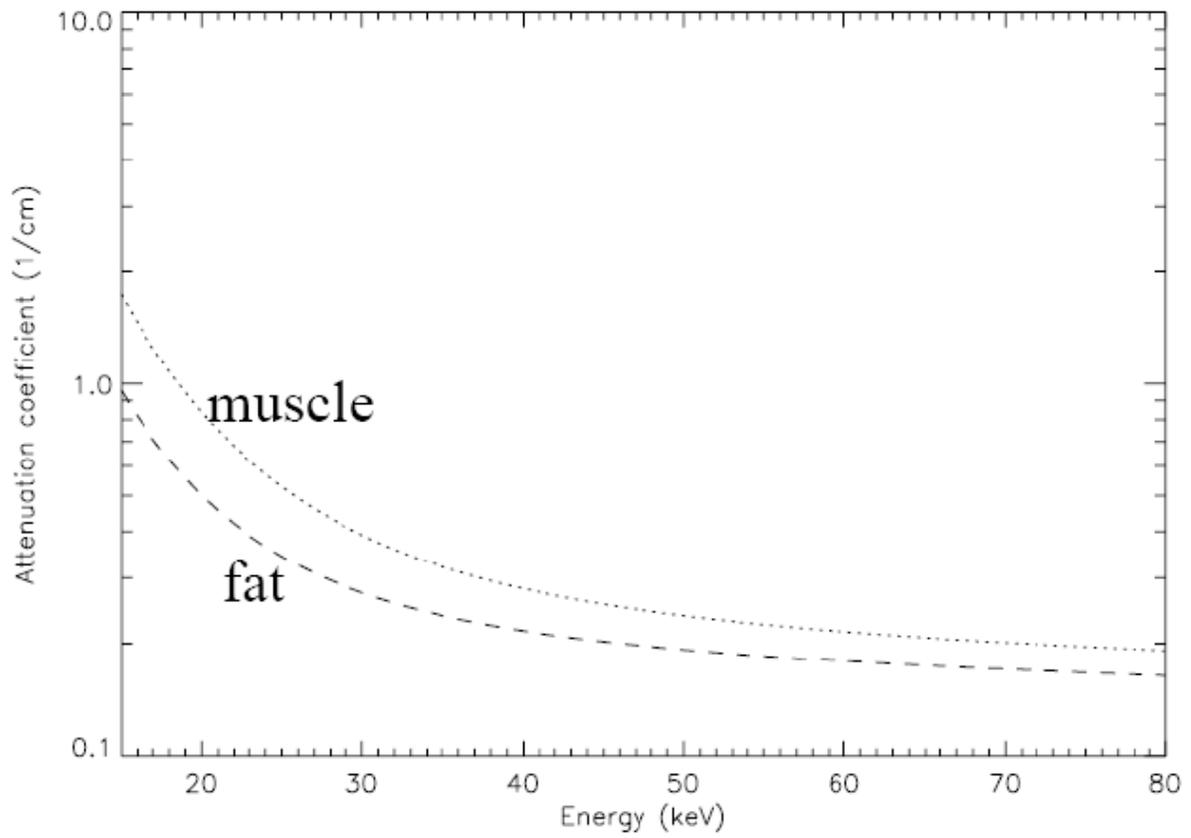

**Figure 2. x-ray linear attenuation coefficient for fat and muscle as function of x-ray energy.**

The chemical composition of fat and muscle assumed for Figure 2 is listed in table 1 [].

| Composition | Fat (mass percentage) | Muscle (mass percentage) |
|---|---|---|
| Hydrogen | 11.2% | 10.2% |
| Carbon | 57.3% | 12.3% |
| Nitrogen | 1.1% | 3.5% |
| Oxygen | 30.3% | 72.9% |
| Sulfur | .06% | .5% |
| Sodium | | .08% |
| Magnesium | | .02% |
| Phosphour | | .2% |
| Calcium | | .007% |
| Potassium | | .008% |

**Figure 2. Chemical composition of fat and muscle used for theoretical calculation of x-ray linear attenuation coefficients**

Figure 2 shows that the dependence of fat and muscle on the x-ray energy is non-linear is different due to different elemental composition. Thus, by measuring the x-ray transmission ($I_l$ and $I_h$) at two different x-ray energies ($E_l$ and $E_h$), one can solve the thickness of the fat and muscle.

$$I_l = I_{0l} \exp(-\mu_{lf} T_f) \exp(-\mu_{lm} T_m) \qquad \text{Eq. 3}$$

$$I_h = I_{0h}\exp(-\mu_{hf}T_f)\exp(-\mu_{hm}T_m) \qquad \text{Eq. 4}$$

where $\mu_{lf}$ and $\mu_{hf}$ are the linear attenuation coefficients of fat at $E_l$ and $E_h$, respectively, and $\mu_{lm}$ and $\mu_{hm}$ are the linear attenuation coefficients of muscle at $E_l$ and $E_h$, respectively.

Taking the logrithmic of Eqs. 3 and 4 results in two linear equations

$$\ln(I_l/I_{0l}) = -\mu_{lf}T_f - \mu_{lm}T_m$$
$$\ln(I_h/I_{0h}) = -\mu_{hf}T_f - \mu_{hm}T_m \qquad \text{Eq. 5, Eq. 6}$$

$\ln(I_l/I_{0l})$, defined as *a*, can be experimentally measured by performing x-ray attenuation measurement at x-ray energy $E_l$. Similarly, $\ln(I_l/I_{0l})$, defined as *b*, can be measured at x-ray energy $E_h$. With these two measurements and knowing the theoretical linear attenuation coefficients of fat and muscle at these two energies, Eq. 5 and 6 can be solved to obtain the thickness of fat and muscle in the sample:

$$T_f = \frac{-a\mu_{lf} - b\mu_{lm}}{\mu_{lf}\mu_{lm} - \mu_{lf}\mu_{lm}} \qquad \text{Eq. 7}$$

$$T_{fm} = \frac{-a\mu_{lf} - b\mu_{lm}}{\mu_{lf}\mu_{lm} - \mu_{lf}\mu_{lm}} \qquad \text{Eq. 8}$$

**Experimental Setup and Data Acquisition**

An experimental apparatus was designed and constructed to measure x-ray attenuation of samples at two x-ray energies simultaneously. Fig. 1 shows the design of the apparatus consisting of radioactive source providing x-rays at 22.105 and 59.5 keV, an energy-sensitive x-ray detector, a Lead collimator and also serves as a sample support, and a detector mount that allows adjustment of the distance between the source and detector.

The x-ray source is a sealed Americium 241 manufactured by Amersham. The Americium 241 emits 59.5 keV x-rays.  A Silver target is used for the source to provide 22.163 keV Kalpha1 and 21.99 keV K alpha2 characteristic x-rays, with the flux-weighted energy being 22.105 keV. With the silver target, x-rays of 2 energies, 22.105 and 59.5 keV are emitted. The flux of the x-rays are 680 ph/s and 500 ph/s at 22.105 and 59.5 keV, respectively, at the detector position.

The x-ray detector is 2008S by Canberra. The distance adjustment between the source and detector is achieved by X95 (Newport) slide. In the final configuration, the distance is 50 mm.

Fig. 3 shows a picture of the experimental setup.

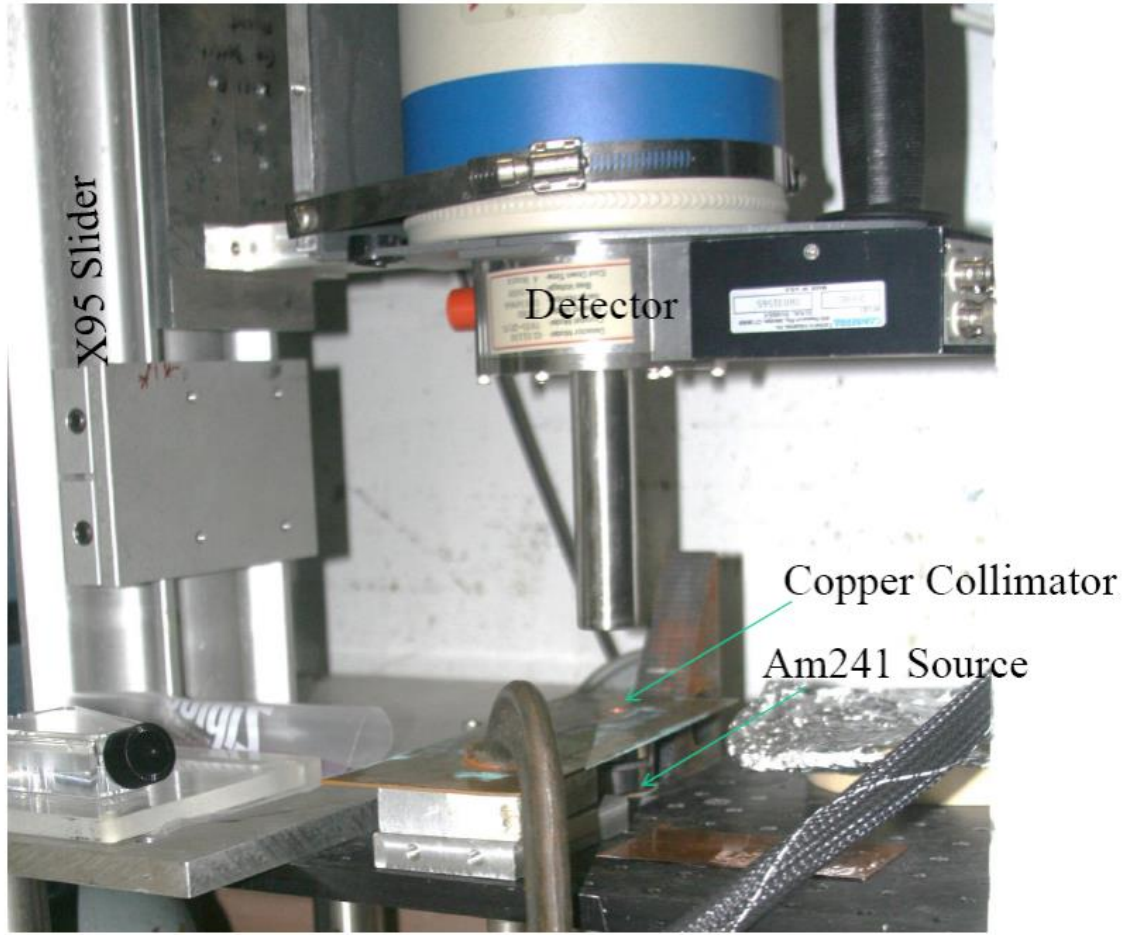

**Fig.3, experimental setup.**

Background spectrum was acquired without the sample for 3600 s (T0) to assure a good quality background. Then the sample was added and a spectrum was acquired for time T which can be different from T0. The background spectrum does not change with time, thus only one background is needed for all the measurements. Figure 4 shows a typical spectrum of the background in solid line, and spectrum with sample in dashed line. It can be seen that the lower-energy x-rays are attenuated more than the high energy x-rays.

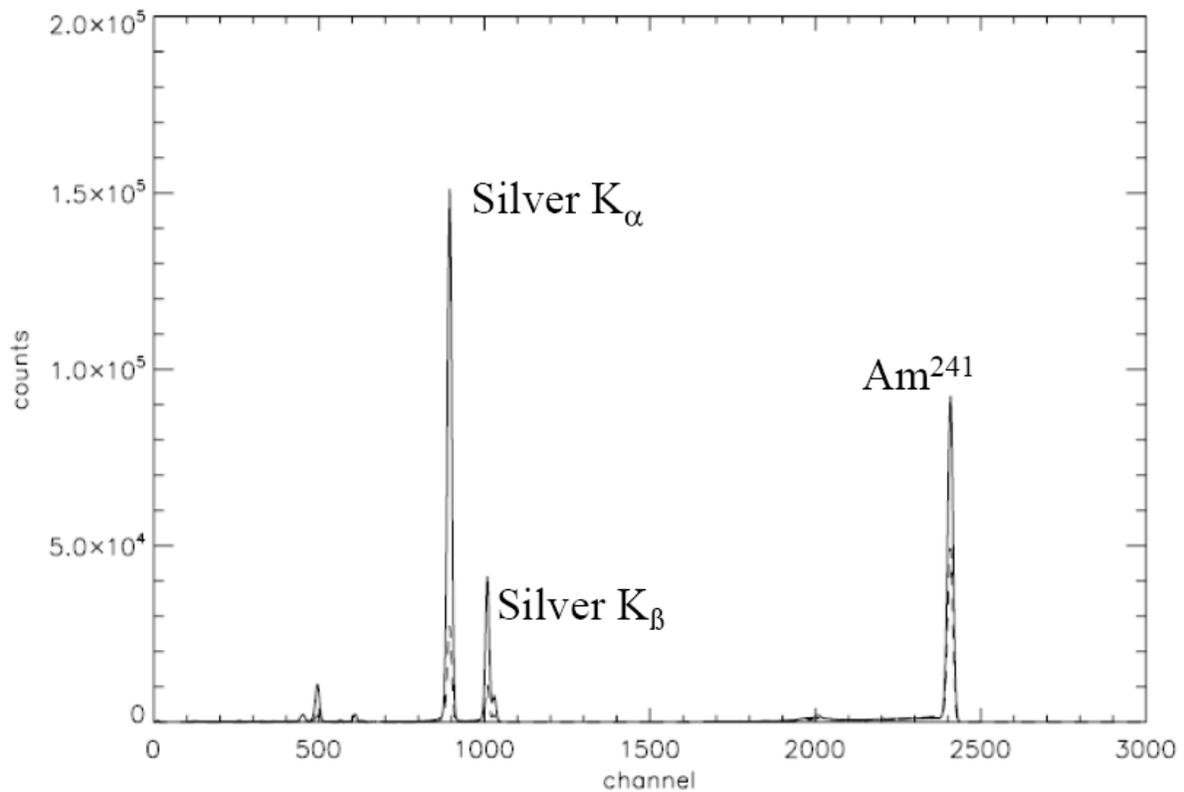

**Figure 4**, Typical raw data spectrum with (dashed line) and without (solid line) the sample showing the detector counts vs. channel number. The channel number is proportional to the x-ray energy.

The $I_0$ and I in Eqs 5 and 6 are derived from experimental data by summarizing the counts under the Silver Ka peak for low energy x-rays, and the Am241 peak for high energy x-rays. Then Eqs. 7 and 8 are used to calculate the thickness of fat and muscle in the sample. An IDL program was developed to perform the data analysis automatically. The program requires user to simply enter the file names for the background and sample spectra, and the acquisition time used for each spectrum.

**Phantom Studies**

Fat and muscle was first simulated using water and plexiglass. The x-ray attenuation coefficients for water, fat, muscle and plexiglass are shown in table 2 for 22.105 and 59.5 keV x-rays. The table shows that the x-ray attenuation of water is similar to that of muscle at both energies, and that the plexiglass is similar to fat. The advantage of using phantom is that we can continuously and accurately vary the thickness of the compositions to study the signal-to-noise ratio of the measurement, and to find out the optimum measurement time.

| Element | $\mu$ at 22.105 keV (cm$^{-1}$) | $\mu$ at 59.5 keV (cm$^{-1}$) |
|---|---|---|
| Fat | 0.4579 | 0.1976 |
| Muscle | 0.6408 | 0.2058 |
| Pexiglas | 0.5456 | 0.2293 |
| Water | 0.6292 | 0.2068 |

**Table 2. Linear attenuation coefficients of fat, muscle, Plexiglas and water shows the similarity between the linear attenuation coefficients of fat and Plexiglas, and between that of muscle and water.**

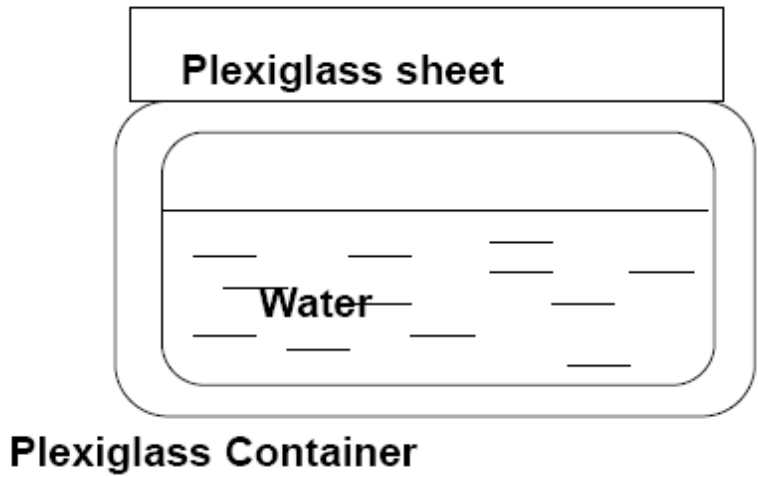

**Figure 5. Design of the phantom used for the study that allows varying plexiglass and water thicknesses.**

Fig. 5 shows the design of the phantom, along with a representative picture of the phantom in one configuration. Adding or removing plexiglass sheets on the top changes the thickness of the plexiglass, adjusting the water level changes the thickness of the water.

Measurements were made with 3 phantoms consisting of 3 mm plexiglass + 9 mm water, 11 mm plexiglass + 9 mm water, and 11 mm plexiglass + 19 mm water. For each phantom, data was taken with acquisition times of 120, 225, 300, 450, 900, 1800 and 3600 s. For each acquisition time, 10-100 independent acquisitions were made and the data was analyzed to yield the water and plexiglass thickness, the water-to-plexiglass ratio. Since the multiple acquisitions were made with each acquisition time, the standard deviation of the thicknesses and ratio were also calculated. The data quality improves with increasing acquisition time. This is evident in the plot of the standard deviation of plexiglass-to-water ratio divided by the ratio, versus acquisition time, shown in Fig. 6. Fig. 6 shows that with 900 s acquisition time, plexiglass to water ratio can be determined with an accuracy of about 5%, which is adequate for specimen measurement. Thus we choose 900 s as the standard acquisition time for subsequent measurements with fat and muscle specimen. The high-quality data of the phantom studies with long acquisition time also shows that the method differentiates the 3 phantoms with good accuracy and linearity.

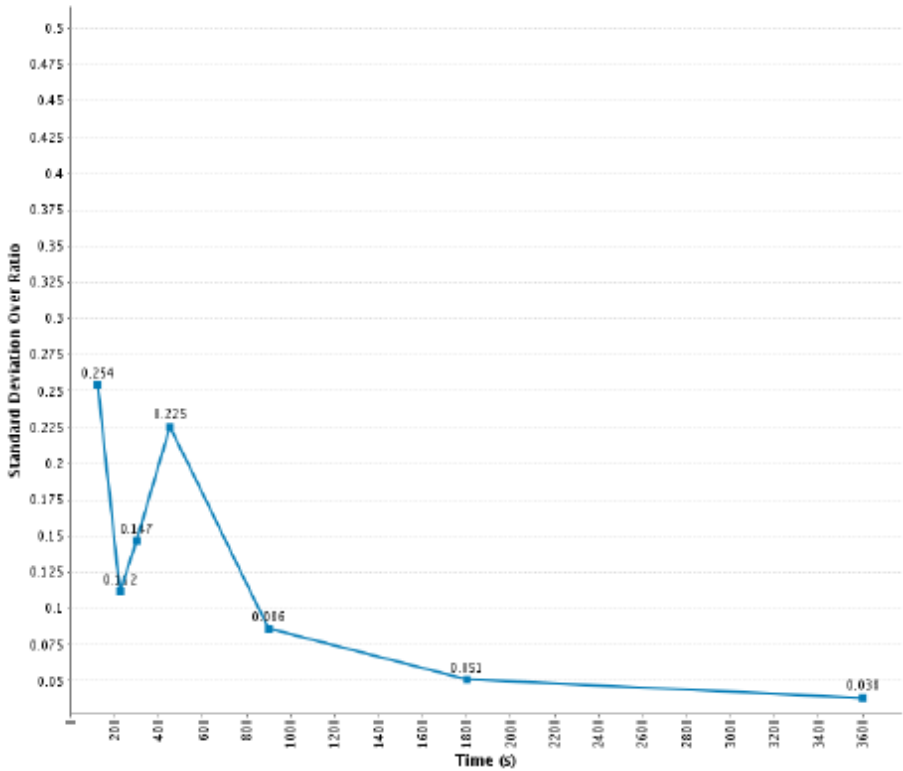

**Figure 6. Accuracy of the plexiglass-to-water ratio versus acquisition time. The accuracy is represented by the standard-deviation of the derived ratio divided by the ratio.**

**Specimen Studies**

Specimen of pork with different fat-to-muscle ratios were prepared by cutting fillets of 5 mm thick by 40 mm square of pure fat and pure muscle. Six fillets are then combined to form specimen of 30 mm total thickness, with 0 fat + 30 mm muscle, 5 mm fat + 25 mm muscle, 10 mm fat + 20 mm muscle and so on. Each 30 mm thick sample was measured with an exposure time of 900s and the thicknesses of fat and muscle were calculated. Figure 7 shows the resulting measured weight (density times measured thickness) for fat (+) and muscle (*) versus theoretical weight. The dashed lines is a guide to the eye for the case when measurement agrees perfectly with theory. Fig. 7 shows that with 900 s measurement time, the method is able to detect fat and muscle with good accuracy and linearity. Note that the 30 mm specimen thickness is similar to the thickness of lab rats and mice.

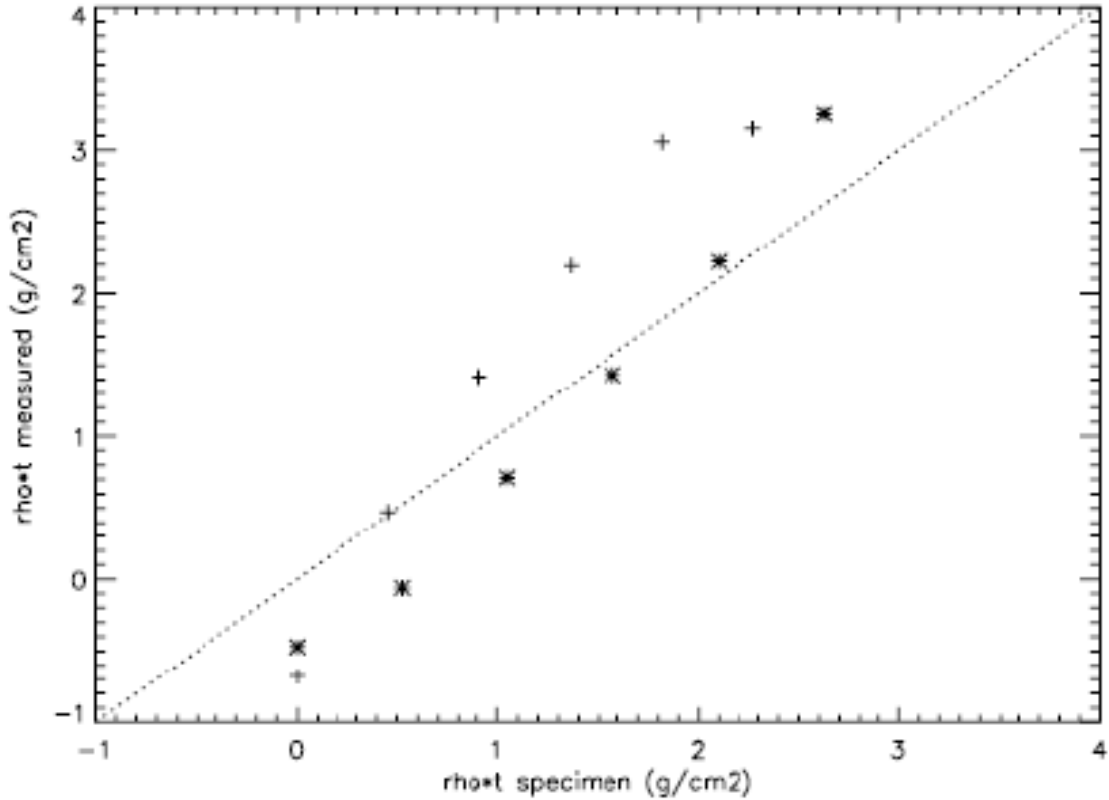

**Figure 7. Measured fat (+) and muscle (*) weight versus true weight of the specimen with 900 s acquisition time.**

With an exposure time of 900s, the incident x-rays flux on the sample are $5 \times 10^5$ ph/s for both x-ray energies. This results in x-ray dose of 0.00084 and 0.00015 mGy for 22.105 and 59.5 keV x-rays. The total x-ray dose is 0.00099 mGy.

**Conclusions and Discussions**

The results of phantoms of plexiglass and water with total thickness of 12 to 40 mm and a variety of Plexiglas-to-water ratio validates the method's utility for measurement in a two-component system with excellent linearity and good sensitivity that improves with longer acquisition time. The study allowed determination of appropriate acquisition time to minimize dosage while obtaining accurate data.

Specimens of pork fat and muscle with different fat to muscle ratios were measured to validate this method's applicability for animal research as well. By using and layering fat and muscle, a condition that is similar to the testing of live animals is created.

The results showed that the method was sensitive to the fat to muscle ratios with good linearity. A standard deviation of a few percent in the fat to muscle ratio could be observed with the x-ray dose of 0.001 mGy. This is compared with the standard micro-CT dose of about 1 mGy. The x-ray source that was used for calibration was fairly weak, and a source that is a hundred times stronger is commercially available. The current experiment uses a measurement time of 900 seconds, however, with a stronger source, the time could be less than 10 seconds. and the cost would be low enough that it will most likely become main-stream in most laboratories. Also, the animals would be allowed to freely move under the x-ray emitter, so anesthetization would be unnecessary.

Further studies with rat specimen and eventually live animals is needed to fully validate the method's utility for obesity research using small animals. Since animal consists of bone in addition to fat and muscle, the method may need to be extended to include 3 energies in order to solve 3 unknown quantities.

**Acknowledgements**

Use of the National Synchrotron Light Source, Brookhaven National Laboratory, was supported by the U.S. Department of Energy, Office of Basic Energy Sciences, under contract number DE-AC02-98CH10886, the Brookhaven National Laboratory LDRD 05-057. We thank Steven Townsend, Lori

Steigler, and Rick Greene (all from Brookhaven Lab) for their expert technical assistance in radiation protection and source management, Nancye Wright and Gretchen Cisco, both from Brookhaven Lab, for their administrative assistance.
**References**

[1] S. Judex, Y.K. Luu, E. Ozcivici, B. Adler, S. Lublinsky, and C. Rubin, Methods, 50 (2010) 14-19.

[2] S. Cummings, D. Bates, D. Black, J. Am. Med. Assoc. **288** (2002) 1889-1897.

[3] E. Rubenstein, J.C. Giacomini, H. J. Gordon, A.C. Thompson, G. Brown, R. Hofstadter, W. Thomlinson, and H.D. Zeman, Nucl. Instrum. Meth. in Phys. Res. A **291** (1990) 80-85.

[4] H. Elleaume, S. Fiedler, F. Esteve, et. al., Phys. Med. Biol. **45** (2000) L49-L43.

S. Ohtsuka, Y. Sugishita, T. Takeda, Y. Itai, J. Tada and K. Hyodo, Br. J. Radiol. **72** (1999) 24–8.

[5 ] Z. Zhong, D. Chapman, W. Thomlinson, F. Arfelli, R. Menk, Nucl. Instrum. Meth. in Phys. Res. A **399** (1997) 480-498.

[6] B. Hasegawa, Physics of Medical X-ray Imaging (1987), chapter 4, Medical Physics Pub. Corp.